\documentclass[aps,showpacs,twocolumn,longbibliography,floatfix]{revtex4-1}
\usepackage{amsfonts}
\usepackage{amssymb}
\usepackage{amsmath}
\usepackage{graphicx}
\usepackage{epsfig}
\usepackage{subfigure}
\usepackage{appendix}
\usepackage{color}
\usepackage{xcolor}
\usepackage{tikz}
\usepackage{hyperref}
\usepackage{clrscode3e}
\usepackage{fullpage}

\hypersetup{hypertex=true,
colorlinks=true,
linkcolor=blue,
anchorcolor=blue,
urlcolor=blue,
citecolor=blue}
\definecolor{orcidlogocol}{HTML}{A6CE39}
\DeclareRobustCommand{\orcidicon}[1]{\href{https://orcid.org/#1}{\tikz[baseline=-0.6ex]\node[fill=orcidlogocol,circle,inner sep=0.25pt,minimum size=1.45ex]{\scriptsize\textcolor{white}{\sffamily iD}};}}
\input{tcilatex}
\begin{document}

\title{Emergent Quantized Zak Phases without Dimerization in a
Multicomponent Fermionic Extended Hubbard Model}
\author{F. X. Liu\,\orcidicon{0009-0008-9468-8972}}
\author{Z. Song\,\orcidicon{0000-0002-3315-4589}}
\email{songtc@nankai.edu.cn}
\affiliation{School of Physics, Nankai University, Tianjin 300071, China}

\begin{abstract}
Quantized Zak phases are commonly associated with dimerized systems, such as
the noninteracting Su--Schrieffer--Heeger (SSH) model. In interacting
systems, however, interactions can induce effective dimerization of particle
clusters even in the absence of explicit dimerization in the underlying
lattice. In this work, we study a one-dimensional multicomponent fermionic
extended Hubbard model under a resonance condition between the on-site and
nearest-neighbor interactions. We find that quantized Zak phases emerge for
odd fermion numbers $n\geqslant 3$, protected by the inversion symmetry of
the effective Hamiltonian. Furthermore, in the presence of a tilted field,
fermion clusters exhibit Bloch oscillations, during which their Zak phases
can be extracted from the evolution of a Gaussian wave packet over one Bloch
period. Our results reveal an interaction-induced mechanism for realizing
quantized Zak phases in multicomponent fermionic systems without explicit
lattice dimerization.
\end{abstract}

\maketitle

\section{Introduction}

\label{Introduction} Topological phases of matter are characterized by
global properties of quantum states that cannot be captured by conventional
symmetry-breaking descriptions. The discovery of the quantization of the
Hall conductance in terms of topological invariants established a
fundamental connection between band topology and observable physical
phenomena \cite{Thouless1982}. Since then, topological concepts have become
central to the study of quantum materials, encompassing topological
insulators, superconductors, and a broad range of engineered quantum systems 
\cite%
{Kitaev2001,Ryu2002,Greiner2002,Murakami2004,Kane2005,Bernevig2006,Fu2007,Fu2007a,Schnyder2008,Ryu2010,Hasan2010,Qi2011,Xu2011,Burkov2011,Young2012,Wang2012,Wang2013,Bardyn2012,Tarruell2012,Lin2014,Xiao2015,Weng2015,Lu2015,Leykam2016,Chiu2016,Kunst2018,Armitage2018}%
. In one dimension, the Zak phase, defined as the Berry phase accumulated
by a Bloch eigenstate across the Brillouin zone, provides a useful
characterization of band topology \cite{Zak1989}. A paradigmatic example is
the Su--Schrieffer--Heeger (SSH) model, in which alternating hopping
amplitudes generate topologically distinct phases and support boundary
states under appropriate boundary conditions \cite{Su1979}. In this setting,
the nontrivial Zak phase is closely associated with the dimerized structure
of the lattice \cite{Rhim2017}.

An important question is whether topological properties can emerge in
systems without explicit lattice dimerization. In particular, interactions
can fundamentally modify the effective dynamics of composite particles,
giving rise to topological features that are absent in the corresponding
single-particle problem. For example, an extended Bose-Hubbard model has
been shown to support topologically nontrivial edge and interface states of
repulsively bound particle pairs, even though its single-particle sector is
topologically trivial \cite{Stepanenko2020}. Related studies have further
explored topological states of interacting particle pairs and their possible
experimental realization \cite{Olekhno2020}. These findings demonstrate that
interactions can generate effective topological structures in the dynamics
of composite objects, thereby extending the conventional single-particle
framework of topological band theory.

The dynamics of interacting particles in optical lattices have also
attracted considerable interest because of advances in ultracold-atom
experiments. Optical lattices provide highly controllable settings for
investigating quantum transport, bound states, and strongly correlated
dynamics \cite{Winkler2006,Foelling2007,Gustavsson2008}. In addition, there
exists another type of bound pair that permits correlated single-particle
tunneling\cite{Jin2011,Lin2014,Zhang2025WSL,Zhang2025Magnetic}. Such a bound pair acts as a
quasi-particle, with an energy band width of the same order as that of a
single particle. In particular, repulsively bound pairs have been observed
experimentally, and their formation and propagation have motivated extensive
theoretical investigations \cite%
{Mahajan2006,Petrosyan2007,Creffield2007,Kuklov2007,Zoellner2008,Wang2008,Valiente2008,Jin2009,Valiente2009,Valiente2010,Javanainen2010,Wang2010,Rosch2008,Zhang2024}
Strong interactions can suppress the dissociation of a particle pair by
energetically separating the bound-pair states from the scattering
continuum. Depending on the microscopic processes, the resulting composite
particles may exhibit either strongly suppressed or appreciable correlated
tunneling, leading to distinct effective energy scales and dynamical
behaviors. These results provide a natural starting point for investigating
whether interaction-induced composite-particle dynamics can also give rise
to quantized geometric phases.

In this work, we investigate a one-dimensional multicomponent fermionic
extended Hubbard model under a resonance condition between the on-site and
nearest-neighbor interactions. We show that the interaction-induced dynamics
of fermion clusters can be described by an effective Hamiltonian whose
inversion symmetry gives rise to quantized Zak phases for odd fermion
numbers $n\geq 3$, even though the underlying lattice does not require
explicit dimerization. This mechanism differs from the conventional SSH
scenario, in which the topological structure originates from alternating
single-particle hopping amplitudes. We further demonstrate that, in the
presence of a tilted field, the fermion clusters undergo Bloch oscillations
and that their Zak phases can be extracted from the evolution of a Gaussian
wave packet over a Bloch period. Our results establish an
interaction-induced route to quantized Zak phases in multicomponent
fermionic systems and suggest a dynamical approach to probing their
topological properties.

The remainder of this paper is organized as follows. Section~\ref{sec:model}
introduces the multicomponent fermionic extended Hubbard model and derives
the effective Hamiltonian under the resonance condition. Section~\ref%
{sec:topology} investigates the inversion symmetry and the quantization of
the Zak phases for fermion clusters. Section~\ref{sec:dynamics} discusses
the Bloch oscillations of fermion clusters and demonstrates how their Zak
phases can be extracted from the wave-packet dynamics. Finally, Sec.~\ref%
{sec_Summary} summarizes our results.

\section{Model and pseudospin representation}

\label{sec:model}

We consider a one-dimensional multicomponent fermionic extended Hubbard
model defined on an $N$-site lattice, with Hamiltonian 
\begin{equation}
H=H_{\mathrm{T}}+H_{\mathrm{V}},  \label{H}
\end{equation}%
where 
\begin{equation}
H_{\mathrm{T}}=-\kappa \sum_{j=1}^{N}\sum_{\alpha =1}^{M}\left( \hat{c}%
_{j,\alpha }^{\dagger }\hat{c}_{j+1,\alpha }+\mathrm{H.c.}\right) ,
\end{equation}%
and%
\begin{equation}
H_{\mathrm{V}}=U\sum_{j=1}^{N}\sum_{\alpha >\beta =1}^{M}\hat{n}_{j,\alpha }%
\hat{n}_{j,\beta }+V\sum_{j=1}^{N}\sum_{\alpha ,\beta =1}^{M}\hat{n}%
_{j,\alpha }\hat{n}_{j+1,\beta },
\end{equation}%
Here, $\hat{c}_{j,\alpha }^{\dagger }$ and $\hat{c}_{j,\alpha }$ denote the
creation and annihilation operators for a fermion of component $\alpha $ at
site $j$, respectively. The component-resolved and total particle-number
operators at site $j$ are defined as 
\begin{equation}
\hat{n}_{j,\alpha }=\hat{c}_{j,\alpha }^{\dagger }\hat{c}_{j,\alpha },\qquad 
\hat{n}_{j}=\sum_{\alpha =1}^{M}\hat{n}_{j,\alpha }.
\end{equation}%
The parameters $\kappa $, $U$, and $V$ represent the tunneling amplitude,
on-site interaction, and nearest-neighbor interaction strength, respectively.

Under periodic boundary conditions, 
\begin{equation}
\hat{c}_{j+N,\alpha }=\hat{c}_{j,\alpha },
\end{equation}%
the Hamiltonian possesses several symmetries and consequently decomposes
into invariant subspaces. For example, 
\begin{equation}
\lbrack \hat{n},H]=0,\qquad \lbrack \hat{T}_{1},H]=0,
\end{equation}%
where 
\begin{equation}
\hat{n}=\sum_{j=1}^{N}\hat{n}_{j}
\end{equation}%
is the total fermion-number operator, and $\hat{T}_{1}$ is the lattice
translation operator defined by 
\begin{equation}
\hat{T}_{1}\hat{c}_{j,\alpha }\hat{T}_{1}^{-1}=\hat{c}_{j+1,\alpha }.
\end{equation}%
These symmetries provide a useful starting point for analyzing the many-body
spectrum.

In the following, we focus on the resonant condition 
\begin{equation}
U=V.
\end{equation}%
To reveal the structure induced by this resonance, we first consider the
interaction Hamiltonian for a nearest-neighbor dimer consisting of sites $j$
and $j+1$. Its interaction energy can be written as 
\begin{equation}
H_{\text{\textrm{V}}}^{(j,j+1)}=\frac{U}{2}\left[ \hat{n}_{j}(\hat{n}_{j}-1)+%
\hat{n}_{j+1}(\hat{n}_{j+1}-1)+2\hat{n}_{j}\hat{n}_{j+1}\right] .
\end{equation}

Under the resonant condition $U=V$, this expression simplifies to 
\begin{equation}
H_{V}^{(j,j+1)}=\frac{U}{2}(\hat{n}_{j}+\hat{n}_{j+1})(\hat{n}_{j}+\hat{n}%
_{j+1}-1).
\end{equation}

Thus, the interaction energy of a nearest-neighbor dimer depends only on its
total particle number, 
\begin{equation}
\hat{n}_{\mathrm{d}}=\hat{n}_{j}+\hat{n}_{j+1},
\end{equation}%
and is independent of how the particles are distributed between the two
sites. Specifically, 
\begin{equation}
H_{V}^{(j,j+1)}=\frac{U}{2}\hat{n}_{\mathrm{d}}(\hat{n}_{\mathrm{d}}-1).
\end{equation}

Consequently, for a fixed dimer particle number $\hat{n}_{\mathrm{d}}=n$,
all configurations with different particle distributions between sites $j$
and $j+1$ are degenerate under the interaction Hamiltonian. This degeneracy
provides a natural basis for introducing a pseudospin description of the
dimer degrees of freedom.

For a given nearest-neighbor dimer, we introduce the operators 
\begin{equation}
\hat{J}_{+}=\left( \hat{J}_{-}\right) ^{\dag }=\sum_{\alpha =1}^{M}\hat{c}%
_{j,\alpha }^{\dagger }\hat{c}_{j+1,\alpha },
\end{equation}%
and 
\begin{equation}
\hat{J}_{z}=\frac{\hat{n}_{j}-\hat{n}_{j+1}}{2}.
\end{equation}

These operators describe the transfer of fermions between the two sites and
the corresponding particle-number imbalance. They satisfy the pseudospin
algebra 
\begin{equation}
\lbrack \hat{J}_{+},\hat{J}_{-}]=2\hat{J}_{z},
\end{equation}%
and%
\begin{equation}
\lbrack \hat{J}_{z},\hat{J}_{\pm }]=\pm \hat{J}_{\pm }.
\end{equation}

\begin{figure*}[t]
\centering
\includegraphics[width=0.90\textwidth,height=0.8%
\textheight,keepaspectratio]{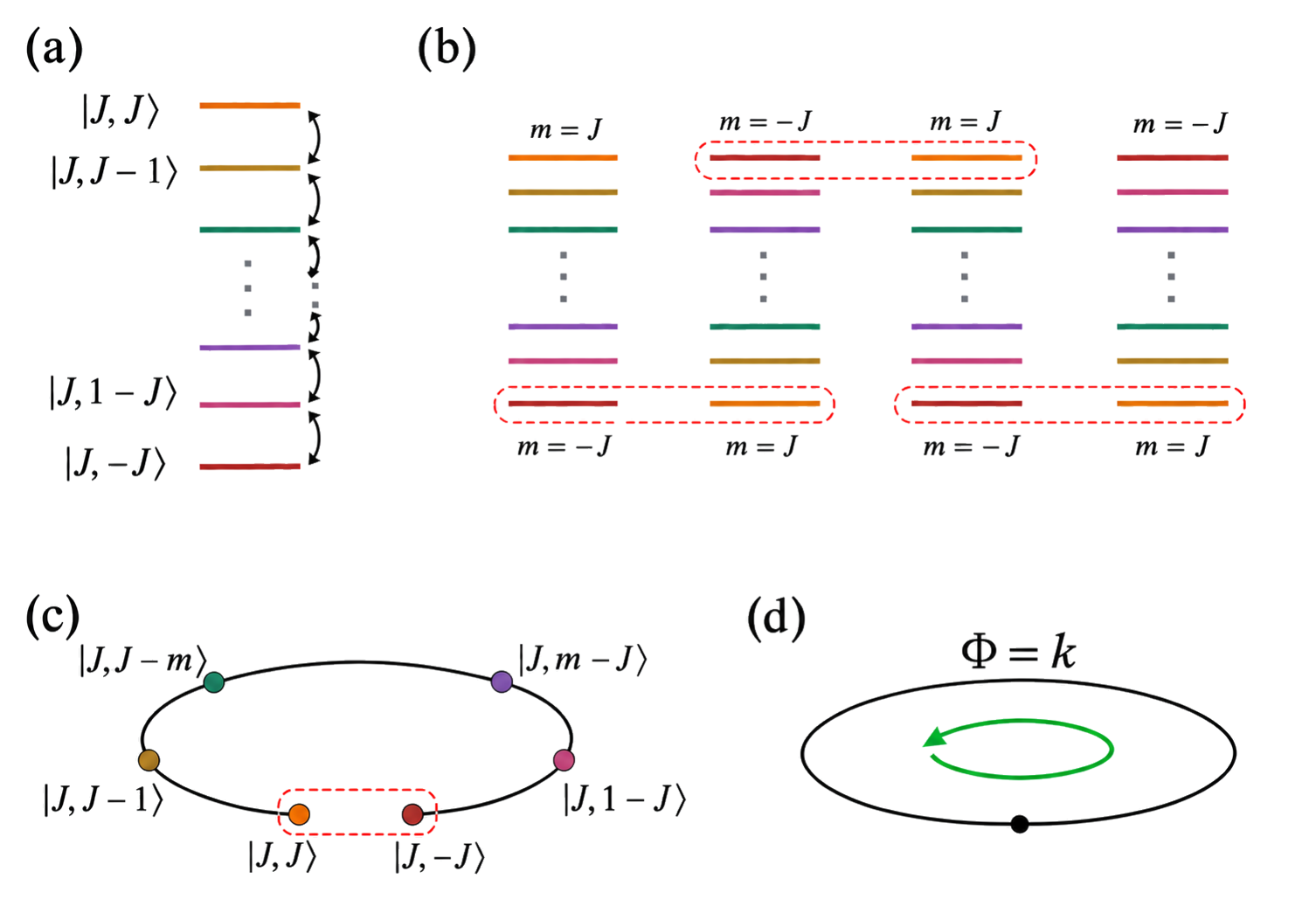}
\caption{Pseudospin representation and construction of the effective
lattice. (a) Pseudospin ladder for the effective dimer Hamiltonian under the
resonant condition. (b) Pseudospin representation of the effective
Hamiltonian for a chain in the $n=2J$ filling sector, obtained by \textit{%
combining} the endpoint configurations $|J,-J\rangle _{j,j+1}=|J,J\rangle
_{j+1,j+2}$ of adjacent dimers. (c) First step in constructing the
momentum-space representation of the effective Hamiltonian in the $k$
sector, obtained by repeating the pseudospin ladder in (a) and replacing $%
|J,m\rangle $ with $|J,m\rangle _{k}$. (d) Second step: \textit{combining}
the two endpoint configurations $|J,-J\rangle _{k}=|J,J\rangle _{k}$ to form
an $n$-site ring. The third step is to introduce a flux phase $\Phi =k$
threading the ring. Different choices of the unit cell correspond to placing
the factor $e^{ik}$ on different bonds without changing the total Peierls
phase. We note that for odd $n$, the ring system possesses inversion
symmetry when the factor $e^{ik}$ is assigned to the hopping amplitude
between the sites $m=1$ and $m=-1$.}
\label{fig1}
\end{figure*}

The dimer Hilbert space can therefore be organized into irreducible
pseudospin multiplets labeled by $J$ and $m$. The corresponding basis states
satisfy 
\begin{eqnarray}
\hat{J}^{2}|J,m\rangle &=&J(J+1)|J,m\rangle ,  \notag \\
\hat{J}_{z}|J,m\rangle &=&m|J,m\rangle ,
\end{eqnarray}%
and 
\begin{equation}
\hat{J}_{\pm }|J,m\rangle =\sqrt{J(J+1)-m(m\pm 1)}\,|J,m\pm 1\rangle .
\end{equation}%
Indeed, for a fixed total dimer particle number $n$, starting from the basis
states%
\begin{equation}
|J,J\rangle =\left\vert \frac{n}{2},\frac{n}{2}\right\rangle
_{j,j+1}=\prod_{\alpha =1}^{n}c_{j,\alpha }^{\dagger }|\mathrm{vac}\rangle ,
\end{equation}%
and%
\begin{equation}
|J,-J\rangle =\left\vert \frac{n}{2},-\frac{n}{2}\right\rangle
_{j,j+1}=\prod_{\beta =1}^{n}c_{j+1,\beta }^{\dagger }|\mathrm{vac}\rangle ,
\end{equation}%
all the $n+1$ states $\left\{ |J,m\rangle \right\} $ can be obtained
explicitly, where $\left\vert \mathrm{vac}\right\rangle $\ is the vacuum
state for all the fermion operator $c_{l,\alpha }$.\ This indicates that the
interaction Hamiltonian is proportional to the identity within each
degenerate pseudospin manifold. The tunneling term, on the other hand, acts
as a pseudospin-flip operator and mixes states within these manifolds. This
pseudospin structure will be useful for constructing the effective
description of the many-body problem and for identifying the special
properties associated with the resonant interaction condition. The
conclusion holds under the filling condition $n\leqslant M$. Fig. \ref{fig1}%
(a) illustrates the pseudospin representation for the effective dimer
Hamiltonian under the resonant condition. 

\section{Topological effective Hamiltonian}

\label{sec:topology}

We now extend the single-dimer result to an $N$-cell lattice. Within the
invariant subspace with a fixed filling number $n$, we introduce a set of
basis states $\{|l\rangle \}$ with $l=1,\ldots ,nN$, defined by 
\begin{equation}
|(j-1)n+\lambda \rangle =\left\vert \frac{n}{2},\frac{n}{2}-\lambda
+1\right\rangle _{j,j+1},
\end{equation}%
where $j=1,\ldots ,N$ labels the unit cells and $\lambda =1,\ldots ,n$
labels the internal states within each unit cell.

In this basis, the effective Hamiltonian takes the form 
\begin{equation}
H_{\mathrm{eff}}^{[n]}=-\kappa \sum_{j}h_{j}-\kappa \sum_{j}I_{j,j+1}+\frac{%
Un(n-1)}{2}\sum_{l=1}^{nN}|l\rangle \langle l|,
\end{equation}%
where 
\begin{eqnarray}
h_{j} &=&\sum_{\lambda =1}^{n-1}\sqrt{(n-\lambda +1)\lambda }%
\,|(j-1)n+\lambda \rangle  \notag \\
&&\times \langle (j-1)n+\lambda +1|+\mathrm{H.c.},
\end{eqnarray}%
and 
\begin{equation}
I_{j,j+1}=\sqrt{n}\,|jn\rangle \langle jn+1|+\mathrm{H.c.}.
\end{equation}

In its present form, $H_{\mathrm{eff}}^{[n]}$ is formally analogous to a
single-particle tight-binding Hamiltonian on a ring with nearest-neighbor
hopping. The effective lattice consists of $N$ unit cells, each containing $%
n $ internal sites described by the sub-Hamiltonian $h_{j}$. The operators $%
I_{j,j+1}$ describe hopping between neighboring unit cells. Fig. \ref{fig1}%
(b) illustrates the pseudospin representation for the effective Hamiltonian
for a chain in the $n=2J$\ filling sector under the resonant condition.

In this work, we focus on the dynamics of fermionic clusters, which can be
understood in terms of the eigenvalue problem of the effective Hamiltonian.
It is important to emphasize that the effective hopping amplitudes are of
order $\kappa$, rather than $\kappa^2/U$. This distinction is essential: the
relevant dynamics occurs at first order in the hopping amplitude and is
therefore directly accessible in experiments, in close analogy with
single-particle quantum dynamics.

Owing to the translational symmetry of the original system, the effective
Hamiltonian is invariant under translations by one unit cell, 
\begin{equation}
T_n H_{\mathrm{eff}}^{[n]} T_n^{-1} = H_{\mathrm{eff}}^{[n]},
\end{equation}
where the translation operator is defined by 
\begin{equation}
T_n|l\rangle=|l+n\rangle.
\end{equation}
Here and below, the site labels are understood modulo $nN$ for periodic
boundary conditions.

This translational symmetry allows the effective Hamiltonian to be
diagonalized using a Fourier transformation. Its spectrum consequently
consists of $n$ energy bands. The corresponding complete set of Bloch
eigenstates spans the entire invariant Hilbert-space sector associated with
the fixed filling number $n$. It is worth noting that the choice of unit
cell is not unique. For a given Hamiltonian $H_{\mathrm{eff}}^{[n]}$, the
translational symmetry allows $n$ inequivalent choices of the unit-cell
origin. Each choice leads to a different Bloch representation and,
consequently, to a different representation of the corresponding Zak phases.
These representations are related by changes of Bloch convention and should
not be interpreted as different physical Hamiltonians.

Indeed, the matrix representation of the effective Hamiltonian $H_{\mathrm{%
eff}}^{[n]}$ in momentum space can be constructed through the following
steps.

(i) We first introduce a set of basis states in the form 
\begin{equation}
\left\{ \left\vert \frac{n}{2},\frac{m}{2}\right\rangle _{k}\right\} ,\qquad
m=n,n-2,\ldots ,-n.
\end{equation}

(ii) In this basis, the nonzero matrix elements corresponding to
nearest-neighbor hopping are 
\begin{align}
& \left\langle \frac{n}{2},\frac{m}{2}\right\vert _{k}H_{\mathrm{eff}%
}^{[n]}\left\vert \frac{n}{2},\frac{m^{\prime }}{2}\right\rangle _{k}  \notag
\\
& \quad =-\kappa \sqrt{\frac{n}{2}\left( \frac{n}{2}+1\right) -\frac{m}{2}%
\left( \frac{m}{2}+1\right) }\,\delta _{m^{\prime },m+2}  \notag \\
& \qquad -\kappa \sqrt{\frac{n}{2}\left( \frac{n}{2}+1\right) -\frac{m}{2}%
\left( \frac{m}{2}-1\right) }\,\delta _{m^{\prime },m-2}.
\end{align}%
Here, $m,m^{\prime }=n,n-2,\ldots ,-n$. The resulting matrix represents an $%
(n+1)$-site chain with nearest-neighbor hopping.

(iii) We then identify the two end states of the chain according to 
\begin{equation}
\left|\frac{n}{2},\frac{n}{2}\right\rangle_k \equiv \left|\frac{n}{2},-\frac{%
n}{2}\right\rangle_k.
\end{equation}
This identification reduces the number of independent sites from $n+1$ to $n$%
, thereby forming an $n$-site ring.

(iv) Finally, we introduce the crystal momentum by attaching a
momentum-dependent phase to one of the hopping amplitudes, 
\begin{equation}
\kappa \longrightarrow \kappa e^{\pm ik}.
\end{equation}%
The position of this phase factor is determined by the chosen unit-cell
convention. In the following, we illustrate our construction using systems
with small values of $n$.

(i) For $n=3$, the three possible Bloch Hamiltonians, corresponding to the
placement of the momentum-dependent phase on each of the three bonds, are
explicitly given by 
\begin{equation}
h_{k}^{(1)}=%
\begin{pmatrix}
0 & \sqrt{3} & 2e^{-ik} \\ 
\sqrt{3} & 0 & \sqrt{3} \\ 
2e^{ik} & \sqrt{3} & 0%
\end{pmatrix}%
,
\end{equation}%
\begin{equation}
h_{k}^{(2)}=%
\begin{pmatrix}
0 & \sqrt{3}e^{ik} & 2 \\ 
\sqrt{3}e^{-ik} & 0 & \sqrt{3} \\ 
2 & \sqrt{3} & 0%
\end{pmatrix}%
,
\end{equation}%
and 
\begin{equation}
h_{k}^{(3)}=%
\begin{pmatrix}
0 & \sqrt{3} & 2 \\ 
\sqrt{3} & 0 & \sqrt{3}e^{ik} \\ 
2 & \sqrt{3}e^{-ik} & 0%
\end{pmatrix}%
.
\end{equation}

\begin{figure*}[t]
\centering
\includegraphics[width=\textwidth]{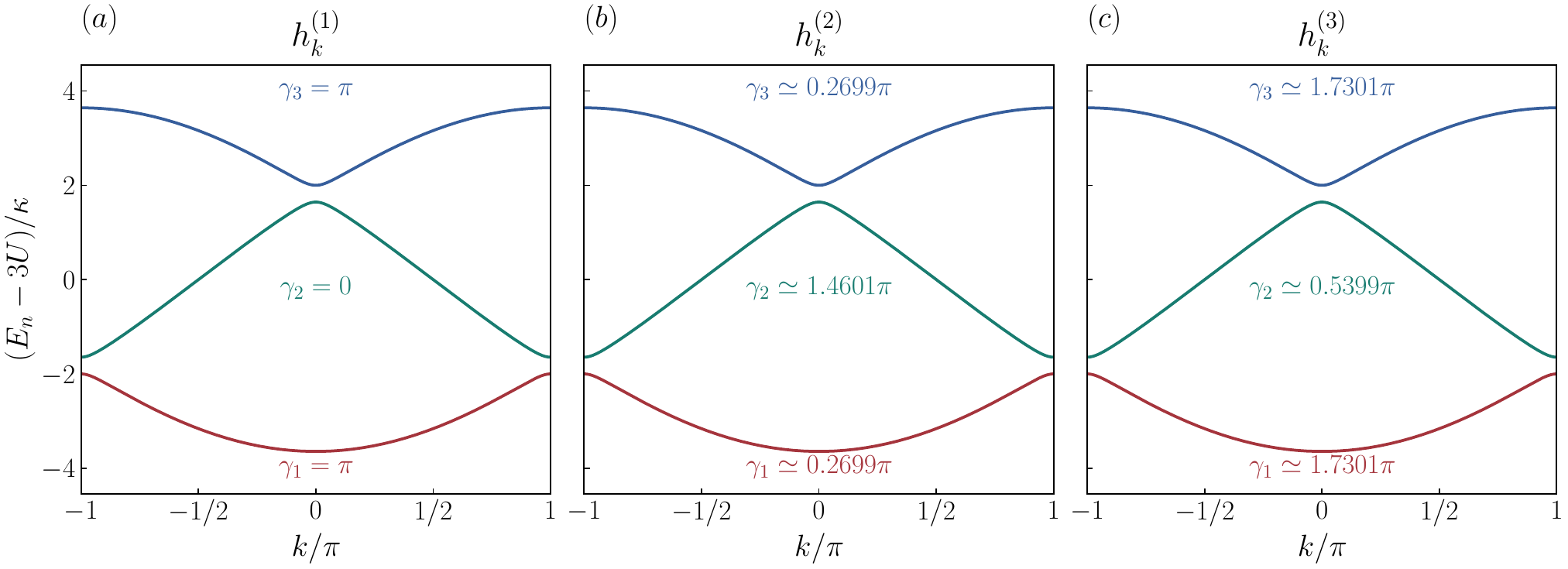}
\caption{Three-particle energy bands and Zak phases for three unit-cell
conventions. Panels (a)--(c) show the eigenvalues of $-h_k^{(\mu)}$,
$\mu=1,2,3$, as $(E_n-3U)/\kappa$, where
$H_{\mathrm{eff}}^{[3]}(k)=3U I-\kappa h_k^{(\mu)}$.
Bands are numbered by increasing physical energy, with their Zak phases
indicated. Inversion symmetry quantizes the phases in panel (a).}
\label{fig:unit-cell-bands}
\end{figure*}

(ii) For $n=4$, the corresponding Bloch Hamiltonian can be written in a
unified form as 
\begin{equation}
h_{k}=%
\begin{pmatrix}
0 & 2\rho _{1} & 0 & 2\rho _{4}^{\ast } \\ 
2\rho _{1}^{\ast } & 0 & \sqrt{6}\rho _{2} & 0 \\ 
0 & \sqrt{6}\rho _{2}^{\ast } & 0 & \sqrt{6}\rho _{3} \\ 
2\rho _{4} & 0 & \sqrt{6}\rho _{3}^{\ast } & 0%
\end{pmatrix}%
,
\end{equation}%
where the phase factors $\rho _{j}$ specify the position of the
momentum-dependent hopping. Specifically, for each choice $j_{0}=1,2,3,4$,
we set 
\begin{equation}
\rho _{j_{0}}=e^{ik},\qquad \rho _{j}=1\quad (j\neq j_{0}).
\end{equation}%
Thus, the four choices of $j_{0}$ correspond to the four possible placements
of the Bloch phase factor on the four bonds of the ring.

(iii) For $n=5$, the corresponding Bloch Hamiltonian can be written in a
unified form as 
\begin{equation}
h_{k}=%
\begin{pmatrix}
0 & 2\sqrt{2}\rho _{1} & 0 & 0 & 3\rho _{5}^{\ast } \\ 
2\sqrt{2}\rho _{1}^{\ast } & 0 & \sqrt{5}\rho _{2} & 0 & 0 \\ 
0 & \sqrt{5}\rho _{2}^{\ast } & 0 & \sqrt{5}\rho _{3} & 0 \\ 
0 & 0 & \sqrt{5}\rho _{3}^{\ast } & 0 & 2\sqrt{2}\rho _{4} \\ 
3\rho _{5} & 0 & 0 & 2\sqrt{2}\rho _{4}^{\ast } & 0%
\end{pmatrix}%
,
\end{equation}%
where the phase factors $\rho _{j}$ specify the position of the
momentum-dependent hopping. Specifically, for each choice $j_{0}=1,2,3,4,\text{ and }5$,
we set 
\begin{equation}
\rho _{j_{0}}=e^{ik},\qquad \rho _{j}=1\quad (j\neq j_{0}).
\end{equation}%
Although the original system considered in Ref.~\cite{Zhang2025Slinky} is a bosonic model, the effective Bloch
Hamiltonians obtained here have exactly the same matrix representations.
Therefore, the corresponding Zak phases and edge modes are identical to
those reported previously, and we do not repeat the detailed topological
analysis. Fig. \ref{fig1}(c) and (d) illustrate the construction of the
pseudospin representation of the effective Hamiltonian in momentum space.

\section{Dynamical signatures of many-body topology}

\label{sec:dynamics}

In this section, we investigate the dynamical signatures of many-body
topology via Bloch oscillations. Here, we focus on the case of odd $n$ for
the following reasons. For even $n$, two energy bands are degenerate, which
makes the Zak phase of an individual band ill-defined. In contrast, for odd $%
n$, an appropriate choice of the unit-cell convention allows one of the
corresponding Bloch Hamiltonians $h_{k}$ to possess inversion symmetry. As a
result, the Zak phases of its isolated bands are quantized to $0$ or $\pi $
modulo $2\pi $.

\begin{figure*}[!t]
\centering
\makebox[\textwidth][c]{\includegraphics[width=1.05%
\textwidth]{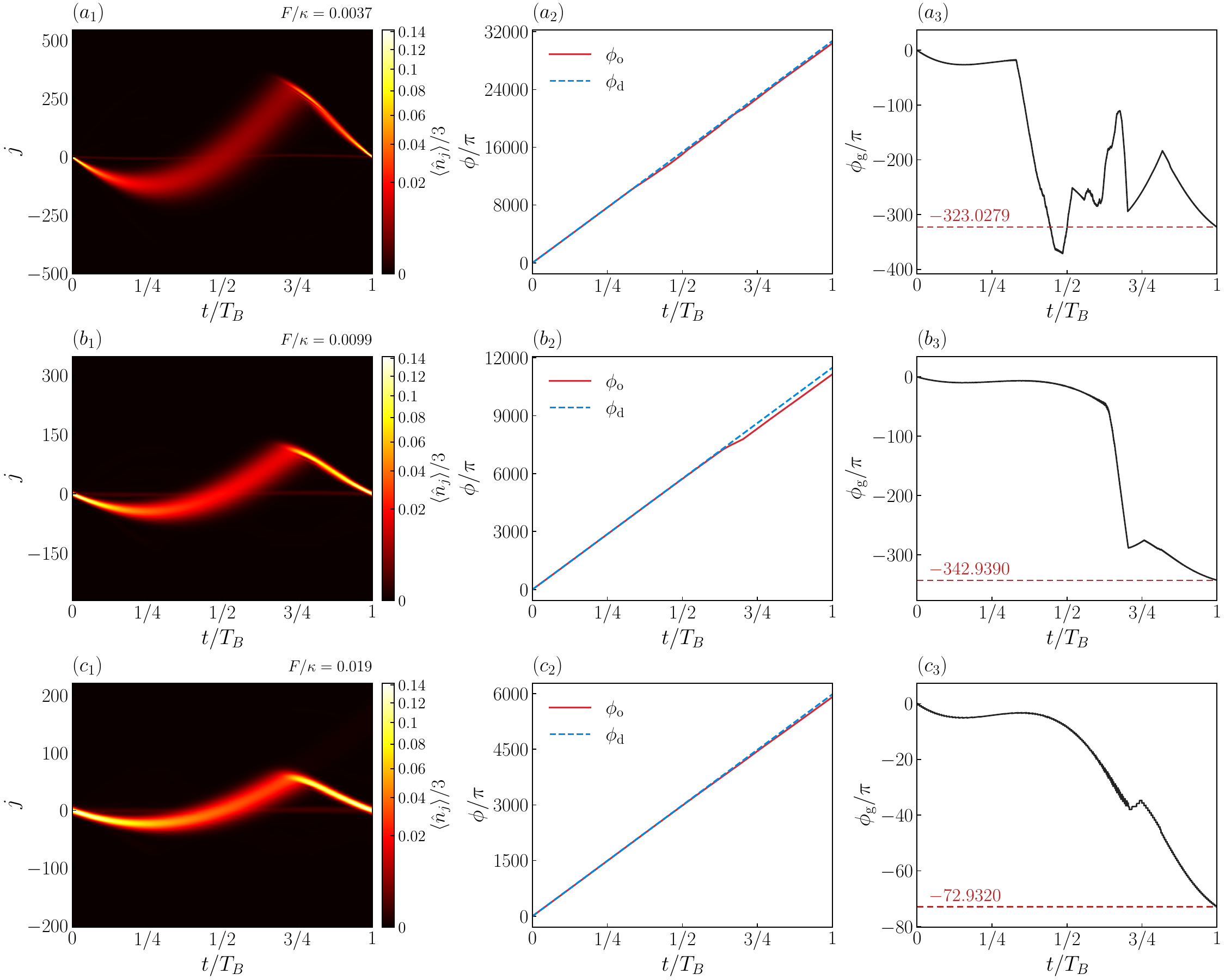}}
\caption{Three-particle Bloch dynamics for the initial state in
Eq.~\eqref{pi}. Columns show $\langle\hat n_j\rangle/3$,
$(\phi_{\mathrm{o}},\phi_{\mathrm{d}})$, and $\phi_{\mathrm{g}}$.
Here $\kappa=1$, $U=V=20$, and $\sigma_k=0.18$.
Rows (a)--(c) use $F/\kappa=0.0037$, $0.0099$, and $0.019$,
respectively. The absolute terminal phase deviations from $\pi$ modulo $2\pi$
are $0.0279\pi$, $0.0610\pi$, and $0.0680\pi$.}
\label{fig:bloch-zak-pi}
\end{figure*}

\begin{figure*}[!t]
\centering
\makebox[\textwidth][c]{\includegraphics[width=1.05%
\textwidth]{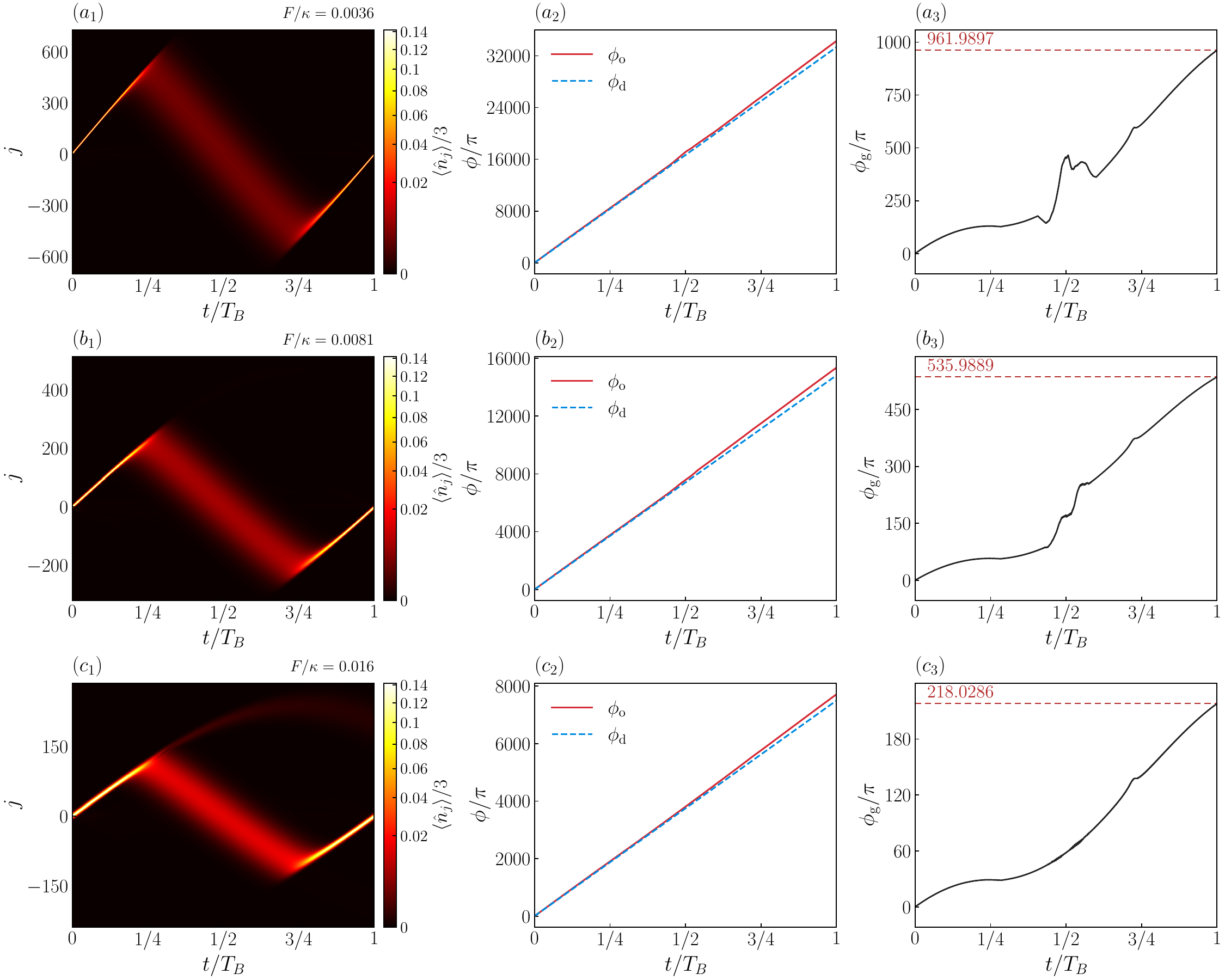}}
\caption{Same quantities as in Fig.~\ref{fig:bloch-zak-pi},
for the initial state in Eq.~\eqref{zero}, with $\kappa=1$, $U=V=20$,
and $\sigma_k=0.18$. Rows (a)--(c) use
$F/\kappa=0.0036$, $0.0081$, and $0.016$, respectively.
The absolute terminal phase deviations from $0$ modulo $2\pi$ are
$0.0103\pi$, $0.0111\pi$, and $0.0286\pi$.}
\label{fig:bloch-zak-zero}
\end{figure*}

Indeed, for odd $n$, the Bloch Hamiltonian can be chosen to possess
inversion symmetry by assigning the momentum-dependent phase to the hopping
between the states with $m=1$ and $m=-1$, 
\begin{eqnarray}
&&\left\langle \frac{n}{2},\frac{1}{2}\right\vert _{k}H_{\mathrm{eff}%
}^{[n]}\left\vert \frac{n}{2},-\frac{1}{2}\right\rangle _{k}  \notag \\
&\longrightarrow &e^{ik}\left\langle \frac{n}{2},\frac{1}{2}\right\vert
_{k}H_{\mathrm{eff}}^{[n]}\left\vert \frac{n}{2},-\frac{1}{2}\right\rangle
_{k}.
\end{eqnarray}%
The resulting Bloch Hamiltonian $h_{k}^{\text{\textrm{In}}}$ satisfies 
\begin{equation}
\mathcal{P}h_{k}^{\text{\textrm{In}}}\mathcal{P}^{-1}=h_{-k}^{\text{\textrm{%
In}}},
\end{equation}%
where the inversion operator $\mathcal{P}$ is defined by 
\begin{equation}
\mathcal{P}\left\vert \frac{n}{2},\frac{m}{2}\right\rangle _{k}=\left\vert 
\frac{n}{2},-\frac{m}{2}\right\rangle _{-k}.
\end{equation}%
Consequently, the Zak phases of the isolated bands are quantized to $0$ or $%
\pi $ modulo $2\pi $.

The topological properties of the effective many-body Hamiltonian can be
probed dynamically through Bloch oscillations. For a given Bloch Hamiltonian 
$h_{k}$, we consider an initial wave packet constructed from an isolated
band $n$, 
\begin{equation}
|\Psi _{n}(0)\rangle =\sum_{k}f(k-k_{0})|u_{n}(k)\rangle ,
\end{equation}%
where $|u_{n}(k)\rangle $ is the Bloch eigenstate of $h_{k}$, and $%
f(k-k_{0}) $ is a narrow Gaussian distribution centered at $k_{0}$.

A constant external force $F$ induces a uniform motion of the quasimomentum
according to the semiclassical equation $\dot{k}=F$, so that 
\begin{equation}
k(t)=k_{0}+Ft.
\end{equation}%
Consequently, the wave packet traverses the entire Brillouin zone during one
Bloch period, 
\begin{equation}
T_{B}=\frac{2\pi }{F}.
\end{equation}

Provided that the evolution remains adiabatic and the wave packet stays
within the same isolated band, its state acquires both a dynamical phase and
a geometric phase. After one complete Bloch cycle, the accumulated geometric
phase is the Zak phase, 
\begin{equation}
\gamma _{n}=i\int_{\mathrm{BZ}}dk\,\langle u_{n}(k)|\partial
_{k}u_{n}(k)\rangle .
\end{equation}%
Thus, the Bloch oscillation provides a dynamical route to probe the
many-body topology encoded in the effective Bloch Hamiltonian $h_{k}$. In
particular, when inversion symmetry is present, the Zak phase is quantized
to $0$ or $\pi $ modulo $2\pi $, and the corresponding distinction between
topologically trivial and nontrivial sectors can be revealed from the
accumulated geometric phase. To extract the geometric contribution, the
total phase accumulated during a Bloch cycle is compared with the
corresponding dynamical phase. The remaining phase is the Zak phase~\cite%
{Atala2013,Ramasesh2017}.

For the following three-particle dynamics, the force is implemented by a
linear potential on the physical lattice sites, 
\begin{equation}
H=H_{T}+H_{V}+H_{F},\qquad H_{F}=-\frac{F}{3}\sum_{j}j\hat{n}_{j},
\label{H_F}
\end{equation}%
where $\hat{n}_{j}=\sum_{\alpha =1}^{3}\hat{n}_{j,\alpha }$ is the total
occupation of site $j$.

Our aim is to investigate the dynamical process of the original many-body
system governed by the Hamiltonian $H$ in Eq.~\eqref{H_F}. To this end, we
construct the initial state from the eigenstates of the effective
Hamiltonian $h_{k}^{\mathrm{In}}$ and subsequently evolve it under the full
Hamiltonian $H$. Although the initial state belongs to the energy shell $%
Un(n-1)/2$, the full Hamiltonian generally couples this subspace to other
many-body sectors. Consequently, the evolved state is not restricted to the
basis states within the original energy shell.

\raggedbottom
\makeatletter
\@booleantrue\raggedcolumn@sw
\makeatother
\begin{samepage}
\postdisplaypenalty=0
As a demonstration, we examine the dynamic process of the Hamiltonian $H$\
with several representative parameters\ on a finite chain in $n=3$\
sector. The initial state has the explicit form%
\begin{equation}
\begin{aligned}
\lvert\psi(0)\rangle={}&\mathcal{N}\sum_j
\exp\!\left[-\frac{(j-j_0)^2}{4\sigma^2}\right] \\
&{}\times e^{-i\pi j/2}\lvert\psi_j^{\pi,0}\rangle,
\end{aligned}
\end{equation}%
\end{samepage}
where $\mathcal{N}$ is a normalization factor. Here
{\predisplaypenalty=0\relax
\begin{eqnarray}
&&\left\vert \psi _{j}^{\pi }\right\rangle =\sqrt{\frac{3}{10}}%
\{c_{j,1}^{\dagger }c_{j,2}^{\dagger }c_{j,3}^{\dagger } \notag \\
&&+\frac{\sqrt{10}-2i}{6}\lbrack c_{j+1,1}^{\dagger }c_{j,2}^{\dagger }c_{j,3}^{\dagger }
+c_{j,1}^{\dagger }c_{j+1,2}^{\dagger }c_{j,3}^{\dagger } \notag \\
&&+c_{j,1}^{\dagger }c_{j,2}^{\dagger }c_{j+1,3}^{\dagger }] \notag \\
&&+\frac{\sqrt{10}+2i}{6}\lbrack c_{j-1,1}^{\dagger }c_{j,2}^{\dagger }c_{j,3}^{\dagger }
+c_{j,1}^{\dagger }c_{j-1,2}^{\dagger }c_{j,3}^{\dagger } \notag \\
&&+c_{j,1}^{\dagger }c_{j,2}^{\dagger }c_{j-1,3}^{\dagger }]\}|\mathrm{vac}%
\rangle , \label{pi}
\end{eqnarray}
}%
and 
\begin{eqnarray}
&&\left\vert \psi _{j}^{0}\right\rangle =\sqrt{\frac{2}{5}}%
\{c_{j,1}^{\dagger }c_{j,2}^{\dagger }c_{j,3}^{\dagger }+\frac{i}{2}%
[(c_{j+1,1}^{\dagger }c_{j,2}^{\dagger }c_{j,3}^{\dagger }  \notag \\
&&+c_{j,1}^{\dagger }c_{j+1,2}^{\dagger }c_{j,3}^{\dagger }+c_{j,1}^{\dagger
}c_{j,2}^{\dagger }c_{j+1,3}^{\dagger })-(c_{j-1,1}^{\dagger
}c_{j,2}^{\dagger }c_{j,3}^{\dagger }  \notag \\
&&+c_{j,1}^{\dagger }c_{j-1,2}^{\dagger }c_{j,3}^{\dagger }+c_{j,1}^{\dagger
}c_{j,2}^{\dagger }c_{j-1,3}^{\dagger })]\}|\mathrm{vac}\rangle ,
\label{zero}
\end{eqnarray}%
are the unit cell states for the upper and middle bands of $h_{k}^{(1)}$,
respectively, which can be obtained by exact diagonalization of the matrix $%
h_{k}^{(1)}$. In the large $U=V$ limit, the corresponding Zak phase is $\pi $
or $0$. The evolved state $\left\vert \psi (t)\right\rangle
=e^{-iHt}\left\vert \psi (0)\right\rangle $\ is computed numerically. We
calculate the density distribution 
\begin{equation}
\langle \hat{n}_{j}\rangle =\left\langle \psi (t)\right\vert \hat{n}%
_{j}\left\vert \psi (t)\right\rangle ,  \label{n_j}
\end{equation}%
the dynamic phase 
\begin{equation}
\phi _{\text{d}}(t)=\int_{0}^{t}\left\langle \psi (t^{\prime })\right\vert
H\left\vert \psi (t^{\prime })\right\rangle \text{d}t^{\prime },
\label{phi_d}
\end{equation}%
and the overall phase%
\begin{equation}
\phi _{\text{o}}(t)=\arg \left\langle \psi (t)\right\vert l_{m}(t)\rangle
-\arg \left\langle \psi (0)\right\vert l_{m}(0)\rangle ,  \label{phi_o}
\end{equation}%
where $\left\vert l_{m}\right\rangle $\ is the ket with maximal $\left\vert
\left\langle \psi (t)\right\vert l_{m}\rangle \right\vert $. The geometric
phase is obtained by%
\begin{equation}
\phi _{\text{g}}(T_{B})=\phi _{\text{o}}(T_{B})-\phi _{\text{d}}(T_{B}).
\label{phi_g}
\end{equation}%
In Figs.~\ref{fig:bloch-zak-pi} and~\ref{fig:bloch-zak-zero}, we plot the
quantities $\langle \hat{n}_{j}\rangle $, $\phi _{\text{g}}(t)$, $\phi _{%
\text{o}}(t)$, and $\phi _{\text{d}}(t)$, for several representative cases.
The terminal phase differences for these field strengths are marked in the
right panels.

\newpage
\section{Summary}

\label{sec_Summary}

In summary, we have investigated the topological properties and dynamical
signatures of fermion clusters in a one-dimensional multicomponent fermionic
extended Hubbard model under the resonance condition $U=V$. By exploiting
the structure of the resonant interaction, we have constructed an effective
Hamiltonian that describes the cluster dynamics in terms of an effective
lattice with multiple internal states per unit cell. This mapping reveals an
interaction-induced route to topological band structures without explicit
dimerization of the underlying lattice. We have shown that, for odd fermion
numbers $n\geq 3$, an appropriate choice of the unit-cell convention allows
the effective Bloch Hamiltonian to possess inversion symmetry, which
quantizes the Zak phases of its isolated bands to $0$ or $\pi $ modulo $2\pi 
$. We have further explored the dynamical signatures of these topological
properties in the presence of a tilted field. The resulting Bloch
oscillations enable the geometric phase accumulated by a Gaussian wave
packet over a Bloch cycle to be extracted by separating it from the
dynamical phase. Our results establish a mechanism for interaction-induced
Zak-phase quantization in multicomponent fermionic systems and demonstrate
the potential of Bloch-oscillation dynamics as a probe of their effective
band topology.

\acknowledgments This work was supported by National Natural Science
Foundation of China (under Grant No. 12374461).

\bibliographystyle{apsrev4-1}
\bibliography{references}

@article{Thouless1982,
  author = {Thouless, D. J. and Kohmoto, M. and Nightingale, M. P. and den Nijs, M.},
  journal = {Phys. Rev. Lett.},
  title = {{Quantized Hall Conductance in a Two-Dimensional Periodic Potential}},
  year = {1982},
  pages = {405--408},
  volume = {49},
  doi = {10.1103/PhysRevLett.49.405}
}

@article{Kitaev2001,
  author = {Kitaev, A Yu},
  journal = {Phys. Usp.},
  title = {{Unpaired Majorana fermions in quantum wires}},
  year = {2001},
  pages = {131},
  volume = {44},
  doi = {10.1070/1063-7869/44/10s/s29}
}

@article{Ryu2002,
  author = {S. Ryu and Y. Hatsugai},
  title = {{Topological origin of zero-energy edge states in particle-hole symmetric systems}},
  journal = {Phys. Rev. Lett.},
  volume = {89},
  pages = {077002},
  year = {2002}
}

@article{Greiner2002,
  author = {M. Greiner and O. Mandel and T. Esslinger and T. W. H{\"a}nsch and I. Bloch},
  title = {{Quantum phase transition from a superfluid to a Mott insulator in a gas of ultracold atoms}},
  journal = {Nature},
  volume = {415},
  pages = {39},
  year = {2002}
}

@article{Murakami2004,
  author = {S. Murakami and N. Nagaosa and S.-C. Zhang},
  title = {{Spin-Hall insulator}},
  journal = {Phys. Rev. Lett.},
  volume = {93},
  pages = {156804},
  year = {2004}
}

@article{Kane2005,
  author = {Kane, C. L. and Mele, E. J.},
  journal = {Phys. Rev. Lett.},
  title = {{Quantum Spin Hall Effect in Graphene}},
  year = {2005},
  pages = {226801},
  volume = {95},
  doi = {10.1103/PhysRevLett.95.226801}
}

@article{Bernevig2006,
  author = {Bernevig, B Andrei and Hughes, Taylor L and Zhang, Shou-Cheng},
  journal = {Science},
  title = {{Quantum spin Hall effect and topological phase transition in HgTe quantum wells}},
  year = {2006},
  pages = {1757--1761},
  volume = {314}
}

@article{Fu2007,
  author = {L. Fu and C. L. Kane},
  title = {{Topological insulators with inversion symmetry}},
  journal = {Phys. Rev. B},
  volume = {76},
  pages = {045302},
  year = {2007}
}

@article{Fu2007a,
  author = {L. Fu and C. L. Kane and E. J. Mele},
  title = {{Topological insulators in three dimensions}},
  journal = {Phys. Rev. Lett.},
  volume = {98},
  pages = {106803},
  year = {2007}
}

@article{Schnyder2008,
  author = {Schnyder, Andreas P. and Ryu, Shinsei and Furusaki, Akira and Ludwig, Andreas W. W.},
  journal = {Phys. Rev. B},
  title = {{Classification of topological insulators and superconductors in three spatial dimensions}},
  year = {2008},
  pages = {195125},
  volume = {78},
  doi = {10.1103/PhysRevB.78.195125}
}

@article{Ryu2010,
  author = {S. Ryu and A. P. Schnyder and A. Furusaki and A. W. W. Ludwig},
  title = {{Topological insulators and superconductors: Tenfold way and dimensional hierarchy}},
  journal = {New J. Phys.},
  volume = {12},
  pages = {065010},
  year = {2010}
}

@article{Hasan2010,
  author = {Hasan, M. Z. and Kane, C. L.},
  journal = {Rev. Mod. Phys.},
  title = {{Colloquium: Topological insulators}},
  year = {2010},
  pages = {3045--3067},
  volume = {82},
  doi = {10.1103/RevModPhys.82.3045}
}

@article{Qi2011,
  author = {X.-L. Qi and S.-C. Zhang},
  title = {{Topological insulators and superconductors}},
  journal = {Rev. Mod. Phys.},
  volume = {83},
  pages = {1057},
  year = {2011}
}

@article{Xu2011,
  author = {G. Xu and H. Weng and Z. Wang and X. Dai and Z. Fang},
  title = {{Chern semimetal and the quantized anomalous Hall effect in {HgCr$_2$Se$_4$}}},
  journal = {Phys. Rev. Lett.},
  volume = {107},
  pages = {186806},
  year = {2011}
}

@article{Burkov2011,
  author = {A. A. Burkov and L. Balents},
  title = {{Weyl semimetal in a topological insulator multilayer}},
  journal = {Phys. Rev. Lett.},
  volume = {107},
  pages = {127205},
  year = {2011}
}

@article{Young2012,
  author = {Young, S. M. and Zaheer, S. and Teo, J. C. Y. and Kane, C. L. and Mele, E. J. and Rappe, A. M.},
  journal = {Phys. Rev. Lett.},
  title = {{Dirac Semimetal in Three Dimensions}},
  year = {2012},
  pages = {140405},
  volume = {108},
  doi = {10.1103/PhysRevLett.108.140405}
}

@article{Wang2012,
  author = {Z. Wang and Y. Sun and X.-Q. Chen and C. Franchini and G. Xu and H. Weng and X. Dai and Z. Fang},
  title = {{Dirac semimetal and topological phase transitions in {$A_3$Bi ($A=$ Na, K, Rb)}}},
  journal = {Phys. Rev. B},
  volume = {85},
  pages = {195320},
  year = {2012}
}

@article{Wang2013,
  author = {Z. Wang and H. Weng and Q. Wu and X. Dai and Z. Fang},
  title = {{Three-dimensional Dirac semimetal and quantum transport in {Cd$_3$As$_2$}}},
  journal = {Phys. Rev. B},
  volume = {88},
  pages = {125427},
  year = {2013}
}

@article{Bardyn2012,
  author = {C.-E. Bardyn and M. A. Baranov and E. Rico and A. {\.{I}mamo{\u g}lu} and P. Zoller and S. Diehl},
  title = {{Majorana modes in driven-dissipative atomic superfluids with a zero Chern number}},
  journal = {Phys. Rev. Lett.},
  volume = {109},
  pages = {130402},
  year = {2012}
}

@article{Tarruell2012,
  author = {L. Tarruell and D. Greif and T. Uehlinger and G. Jotzu and T. Esslinger},
  title = {{Creating, moving and merging Dirac points with a Fermi gas in a tunable honeycomb lattice}},
  journal = {Nature},
  volume = {483},
  pages = {302},
  year = {2012}
}

@article{Lin2014,
  author = {Lin, S. and Zhang, X. Z. and Song, Z.},
  journal = {Phys. Rev. A},
  title = {{Sudden death of particle-pair {B}loch oscillation and unidirectional propagation in a one-dimensional driven optical lattice}},
  year = {2014},
  pages = {063411},
  volume = {90},
  doi = {10.1103/PhysRevA.90.063411}
}

@article{Xiao2015,
  author = {M. Xiao and G. Ma and Z. Yang and P. Sheng and Z. Q. Zhang and C. T. Chan},
  title = {{Geometric phase and band inversion in periodic acoustic systems}},
  journal = {Nat. Phys.},
  volume = {11},
  pages = {240},
  year = {2015}
}

@article{Weng2015,
  author = {Weng, Hongming and Fang, Chen and Fang, Zhong and Bernevig, B. Andrei and Dai, Xi},
  journal = {Phys. Rev. X},
  title = {{Weyl Semimetal Phase in Noncentrosymmetric Transition-Metal Monophosphides}},
  year = {2015},
  pages = {011029},
  volume = {5},
  doi = {10.1103/PhysRevX.5.011029}
}

@article{Lu2015,
  author = {L. Lu and Z. Wang and D. Ye and L. Ran and L. Fu and J. D. Joannopoulos and M. Solja{\v c}i{\'c}},
  title = {{Experimental observation of Weyl points}},
  journal = {Science},
  volume = {349},
  pages = {622},
  year = {2015}
}

@article{Leykam2016,
  author = {D. Leykam and M. C. Rechtsman and Y. D. Chong},
  title = {{Anomalous topological phases and unpaired Dirac cones in photonic Floquet topological insulators}},
  journal = {Phys. Rev. Lett.},
  volume = {117},
  pages = {013902},
  year = {2016}
}

@article{Chiu2016,
  author = {Chiu, Ching-Kai and Teo, Jeffrey C. Y. and Schnyder, Andreas P. and Ryu, Shinsei},
  journal = {Rev. Mod. Phys.},
  title = {{Classification of topological quantum matter with symmetries}},
  year = {2016},
  pages = {035005},
  volume = {88},
  doi = {10.1103/RevModPhys.88.035005}
}

@article{Kunst2018,
  author = {F. K. Kunst and G. van Miert and E. J. Bergholtz},
  title = {{Lattice models with exactly solvable topological hinge and corner states}},
  journal = {Phys. Rev. B},
  volume = {97},
  pages = {241405(R)},
  year = {2018}
}

@article{Armitage2018,
  author = {N. P. Armitage and E. J. Mele and A. Vishwanath},
  title = {{Weyl and Dirac semimetals in three-dimensional solids}},
  journal = {Rev. Mod. Phys.},
  volume = {90},
  pages = {015001},
  year = {2018}
}

@article{Su1979,
  author = {Su, W. P. and Schrieffer, J. R. and Heeger, A. J.},
  journal = {Phys. Rev. Lett.},
  title = {{Solitons in polyacetylene}},
  year = {1979},
  pages = {1698},
  volume = {42},
  doi = {10.1103/PhysRevLett.42.1698}
}

@article{Zak1989,
  author = {Zak, J.},
  journal = {Phys. Rev. Lett.},
  title = {{Berry's phase for energy bands in solids}},
  year = {1989},
  pages = {2747},
  volume = {62},
  doi = {10.1103/PhysRevLett.62.2747}
}

@article{Stepanenko2020,
  author = {A. A. Stepanenko and M. A. Gorlach},
  title = {{Interaction-induced topological states of photon pairs}},
  journal = {Phys. Rev. A},
  volume = {102},
  pages = {013510},
  year = {2020}
}

@article{Winkler2006,
  author = {Winkler, K and Thalhammer, G and Lang, F and Grimm, R and Hecker Denschlag, J and Daley, AJ and Kantian, A and B{\"u}chler, HP and Zoller, P},
  journal = {Nature},
  title = {{Repulsively bound atom pairs in an optical lattice}},
  year = {2006},
  pages = {853--856},
  volume = {441}
}

@article{Foelling2007,
  author = {S. F{\"o}lling and S. Trotzky and P. Cheinet and M. Feld and R. Saers and A. Widera and T. M{\"u}ller and I. Bloch},
  title = {{Direct observation of second-order atom tunnelling}},
  journal = {Nature},
  volume = {448},
  pages = {1029},
  year = {2007}
}

@article{Gustavsson2008,
  author = {Gustavsson, M. and Haller, E. and Mark, M. J. and Danzl, J. G. and Rojas-Kopeinig, G. and N\"agerl, H.-C.},
  journal = {Phys. Rev. Lett.},
  title = {{Control of Interaction-Induced Dephasing of {B}loch Oscillations}},
  year = {2008},
  pages = {080404},
  volume = {100},
  doi = {10.1103/PhysRevLett.100.080404}
}

@article{Mahajan2006,
  author = {S. M. Mahajan and A. Thyagaraja},
  title = {{Exact two-body bound states with Coulomb repulsion in a periodic potential}},
  journal = {J. Phys. A},
  volume = {39},
  pages = {L667},
  year = {2006}
}

@article{Petrosyan2007,
  author = {Petrosyan, David and Schmidt, Bernd and Anglin, James R. and Fleischhauer, Michael},
  journal = {Phys. Rev. A},
  title = {{Quantum liquid of repulsively bound pairs of particles in a lattice}},
  year = {2007},
  pages = {033606},
  volume = {76},
  doi = {10.1103/PhysRevA.76.033606}
}

@article{Creffield2007,
  author = {C. E. Creffield},
  title = {{Coherent control of self-trapping of cold bosonic atoms}},
  journal = {Phys. Rev. A},
  volume = {75},
  pages = {031607(R)},
  year = {2007}
}

@article{Kuklov2007,
  author = {A. Kuklov and H. Moritz},
  title = {{Detecting multiatomic composite states in optical lattices}},
  journal = {Phys. Rev. A},
  volume = {75},
  pages = {013616},
  year = {2007}
}

@article{Zoellner2008,
  author = {S. Z{\"o}llner and H.-D. Meyer and P. Schmelcher},
  title = {{Few-boson dynamics in double wells: From single-atom to correlated pair tunneling}},
  journal = {Phys. Rev. Lett.},
  volume = {100},
  pages = {040401},
  year = {2008}
}

@article{Wang2008,
  author = {L. Wang and Y. Hao and S. Chen},
  title = {{Quantum dynamics of repulsively bound atom pairs in the Bose-Hubbard model}},
  journal = {Eur. Phys. J. D},
  volume = {48},
  pages = {229},
  year = {2008}
}

@article{Valiente2008,
  author = {M. Valiente and D. Petrosyan},
  title = {{Two-particle states in the Hubbard model}},
  journal = {J. Phys. B},
  volume = {41},
  pages = {161002},
  year = {2008}
}

@article{Jin2009,
  author = {Jin, L. and Chen, B. and Song, Z.},
  journal = {Phys. Rev. A},
  title = {{Coherent shift of localized bound pairs in the {B}ose-{H}ubbard model}},
  year = {2009},
  pages = {032108},
  volume = {79},
  doi = {10.1103/PhysRevA.79.032108}
}

@article{Valiente2009,
  author = {M. Valiente and D. Petrosyan},
  title = {{Scattering resonances and two-particle bound states of the extended Hubbard model}},
  journal = {J. Phys. B},
  volume = {42},
  pages = {121001},
  year = {2009}
}

@article{Valiente2010,
  author = {M. Valiente and D. Petrosyan and A. Saenz},
  title = {{Three-body bound states in a lattice}},
  journal = {Phys. Rev. A},
  volume = {81},
  pages = {011601(R)},
  year = {2010}
}

@article{Javanainen2010,
  author = {J. Javanainen and O. Odong and J. C. Sanders},
  title = {{Dimer of two bosons in a one-dimensional optical lattice}},
  journal = {Phys. Rev. A},
  volume = {81},
  pages = {043609},
  year = {2010}
}

@article{Wang2010,
  author = {Y.-M. Wang and J.-Q. Liang},
  title = {{Repulsive bound-atom pairs in an optical lattice with two-body interaction of nearest neighbors}},
  journal = {Phys. Rev. A},
  volume = {81},
  pages = {045601},
  year = {2010}
}

@article{Rosch2008,
  author = {Rosch, Achim and Rasch, David and Binz, Benedikt and Vojta, Matthias},
  journal = {Phys. Rev. Lett.},
  title = {{Metastable Superfluidity of Repulsive Fermionic Atoms in Optical Lattices}},
  year = {2008},
  pages = {265301},
  volume = {101},
  doi = {10.1103/PhysRevLett.101.265301}
}

@article{Zhang2024,
  author = {K.-L. Zhang and X.-D. Jiang and Y.-Y. Li},
  title = {{Two-doublon Bloch oscillations in the mass-imbalanced extended Fermi-Hubbard model}},
  journal = {Phys. Rev. B},
  volume = {110},
  pages = {184304},
  year = {2024}
}

@article{Jin2011,
  author = {Jin, L and Song, Z},
  journal = {New J. Phys.},
  title = {{Fast transfer and efficient coherent separation of a bound cluster in the extended {H}ubbard model}},
  year = {2011},
  pages = {063009},
  volume = {13}
}

@article{Atala2013,
  author = {Marcos Atala and Monika Aidelsburger and Julio T. Barreiro and Dmitry Abanin and Takuya Kitagawa and Eugene Demler and Immanuel Bloch},
  title = {{Direct measurement of the Zak phase in topological Bloch bands}},
  journal = {Nature Physics},
  volume = {9},
  pages = {795--800},
  year = {2013},
  doi = {10.1038/nphys2790}
}

@article{Ramasesh2017,
  author = {Vinay V. Ramasesh and Emmanuel Flurin and Mark S. Rudner and Irfan Siddiqi and Norman Y. Yao},
  title = {{Direct Probe of Topological Invariants Using Bloch Oscillating Quantum Walks}},
  journal = {Phys. Rev. Lett.},
  volume = {118},
  pages = {130501},
  year = {2017},
  doi = {10.1103/PhysRevLett.118.130501}
}

@article{Rhim2017,
  author = {Rhim, Jun-Won and Behrends, Jan and Bardarson, Jens H.},
  title = {{Bulk-boundary correspondence from the intercellular Zak phase}},
  journal = {Phys. Rev. B},
  volume = {95},
  pages = {035421},
  year = {2017},
  doi = {10.1103/PhysRevB.95.035421}
}

@article{Olekhno2020,
  author = {Olekhno, Nikita A. and Kretov, Egor I. and Stepanenko, Andrei A. and Ivanova, Polina A. and Yaroshenko, Vitaly V. and Puhtina, Ekaterina M. and Filonov, Dmitry S. and Cappello, Barbara and Matekovits, Ladislau and Gorlach, Maxim A.},
  title = {{Topological edge states of interacting photon pairs emulated in a topolectrical circuit}},
  journal = {Nat. Commun.},
  volume = {11},
  pages = {1436},
  year = {2020},
  doi = {10.1038/s41467-020-14994-7}
}

@article{Zhang2025WSL,
  author = {Zhang, H. P. and Song, Z.},
  title = {{Formation of generalized Wannier-Stark ladders: Theorem and applications}},
  journal = {Phys. Rev. B},
  volume = {111},
  pages = {014313},
  year = {2025},
  doi = {10.1103/PhysRevB.111.014313}
}

@article{Zhang2025Magnetic,
  author = {Zhang, H. P. and Song, Z.},
  title = {{Bloch oscillations in interacting systems driven by a time-dependent magnetic field}},
  journal = {Phys. Rev. B},
  volume = {111},
  pages = {214306},
  year = {2025},
  doi = {10.1103/1n2v-kxjr}
}

@article{Zhang2025Slinky,
  author = {Zhang, H. P. and Song, Z.},
  title = {{Topological quantum slinky motion in the resonant extended Bose-Hubbard model}},
  journal = {Phys. Rev. B},
  volume = {111},
  pages = {235101},
  year = {2025},
  doi = {10.1103/PhysRevB.111.235101}
}

\end{document}